# Economic Complexity Limits Accuracy of Price Probability Predictions by Gaussian Distributions

Victor Olkhov
Independent, Moscow, Russia
victor.olkhov@gmail.com
ORCID: 0000-0003-0944-5113

April 5, 2024

**Abstract**

We discuss the economic reasons why the predictions of price and return statistical moments in the coming decades, in the best case, will be limited by their averages and volatilities. That limits the accuracy of the forecasts of price and return probabilities by Gaussian distributions. The economic origin of these restrictions lies in the fact that the predictions of the market-based *n*-th statistical moments of price and return for *n=1,2,..,* require the description of the economic variables of the *n-th* order that are determined by sums of the *n*-th degrees of values or volumes of market trades. The lack of existing models that describe the evolution of the economic variables determined by the sums of the 2[nd] degrees of market trades results in the fact that even predictions of the volatilities of price and return are very uncertain. One can ignore existing economic barriers that we highlight but cannot overcome or resolve them. The accuracy of predictions of price and return probabilities substantially determines the reliability of asset pricing models and portfolio theories. The restrictions on the accuracy of predictions of price and return statistical moments reduce the reliability and veracity of modern asset pricing and portfolio theories.

This research received no support, specific grants, or financial assistance from funding agencies in the public, commercial, or nonprofit sectors. We welcome valuable offers of grants, support, and positions.

## 1. Introduction

The asset pricing problem consists of two parts. The first studies asset pricing and portfolio models under the assumption that the predictions of price and return probabilities are known. For example, the consumption-based asset pricing model considers the "mathematical expectation at day *t+1* made by the forecast under the information available at date *t*" (Cochrane, 2001). During the last decades, the asset price theories achieved important results, and references (Markowitz, 1952; Sharpe, 1964; Fama, 1965; Merton, 1973; Fama, 1990; Cochrane, 2001; Fama and French, 2015; Campbell, 2018) present only a millesimal part of the current studies.

The second part of asset pricing forecasts the price and return probabilities, or only their averages and volatilities, at horizon *T*. The price forecasts are studied within economic predictions (Diebold, 1999; Snowberg, Wolfers, and Zitzewitz, 2012), time series analysis (Davis, 1941; Brockwell and Davis, 2002), Monte-Carlo simulations (McLeish, 2005), and, in the last decade, machine learning and AI methods have been applied for stock price predictions (Cao et al., 2021; Kelly and Xiu, 2023). We refer to these papers to highlight the current research, but we have no plans to present a substantive review.

However, we believe that the essential dependence of price and return probabilities on the randomness of market trade hides crucial difficulties of an economic nature, which significantly limit the accuracy of any forecasts of price and return statistical moments by their averages and volatilities and, respectively, reduce the reliability and trustfulness of asset pricing models and portfolio theories.

In this paper, we describe how market-based averages and volatilities of price and return depend on the averages, volatilities, and correlations of the random market trade values and volumes (Olkhov, 2021-2023). We show that such dependence limits the predictability of price and return probabilities. It is well known (Shephard, 1991; Shiryaev, 1999; Shreve, 2004) that a set of statistical moments describes the probability of a random variable. The fewer statistical moments that approximate the probability, the lower the accuracy of the approximation. We consider the economic reasons that limit the number of predicted statistical moments of price and return by their averages and volatilities and that result in a significant decline in the accuracy of any forecasts of their probabilities.

To forecast the averages or volatilities of the price and return of stock of a particular company at horizon *T*, one should predict the averages, volatilities, and correlations of values and volumes of market trade with this stock at the same horizon. That needs the use of the economic and market environment that models trade with stocks of similar companies.

Market trade with a particular stock is always performed by economic agents – a seller and a buyer. We assume that the statistical moments of the values and volumes of market trade with stocks depend on the risk ratings of the traded companies and on ratings of buyers and sellers. To model that dependence, we consider risk grades as numeric continuous risk coordinates. That helps describe the trade with stocks and the statistical moments of the *collective* values and volumes of market trade as functions of risk coordinates. However, the description of the trade statistical moments as functions of risk coordinates requires a model of the evolution of the *joint* values and volumes of the whole market trade. The slow changes of the averages, volatilities, and correlations of the *joint* values and volumes of the whole market trade serve as an economic environment for the description of the statistical moments of the *collective* values and volumes of the market trade.

We highlight that market trade determines economic evolution and the change of macroeconomic variables. The sums of trade values and volumes change agents' variables that we denote 1$^{st}$ order economic variables. In turn, the sums of agents' 1$^{st}$ order variables define macroeconomic variables of the 1$^{st}$ order such as investment, credits, production, etc. (Fox et al., 2017). However, market trade decisions of agents depend on current and future price and return volatilities, which in turn depend on the 2$^{nd}$ statistical moments and correlation of trade values and volume. We show that the sums of squares of trade values and volumes define agents' variables of the 2$^{nd}$ order and they define macroeconomic variables of the 2$^{nd}$ order. Almost each usual variable of the 1$^{st}$ order has its own 2$^{nd}$ order pair. We argue that the predictions of the 2$^{nd}$ statistical moments of trade value and volume, which define price and return volatilities, depend on the joint model of macroeconomic variables of the 1$^{st}$ and the 2$^{nd}$ order that we note as macroeconomic theory of the 2$^{nd}$ order. Such a theory is absent now.

Eventually, our findings are as follows: The current markets provide a lot of trading data that helps "today" approximate the probabilities of the price and return with high accuracy. However, the predictions of the statistical moments at horizon $T$ meet the irremovable barriers of economic complexity. The predictions of price and return volatilities need forecasts of the 2$^{nd}$ statistical moments of market trade. In turn, that requires the economic theory of the 2$^{nd}$ order, which doesn't exist now. In the coming years, in the best scenario, the accuracy of the forecasts will be limited by the 2$^{nd}$ statistical moments of price and return, and hence the forecasts of their probabilities will be limited by Gaussian distributions. The ignorance of the limits driven by economic complexity may allow one to come up with "exact" forecasts of the price and return statistical moments and probabilities.

However, such predictions will have such high uncertainty that it will make them useless or even harmful for investors.

The rest of the paper is organized as follows: In Section 2, we consider the market-based statistical moments of price and return. In Section 3, we consider statistical moments of the *collective* price and return as functions of risk. In Section 4, we consider the market-based averages and volatilities of the *joint* price and return of the whole market. In Section 5, we discuss the origin of the factors that limit the predictions of the statistical moments of price and return. Conclusion in Section 6. In the Appendix, we briefly present the main notions and equations that describe the *collective* trade variables as functions of risk in the continuous economic media approximation.

We are sure that readers know or can find on their own the definitions, terms, and models that are not given in the text. We expect that readers are familiar with common issues in economic theory, asset pricing, risk assessment, probability theory, statistical moments, partial differential equations, etc. Reference (3.5) means equation 5 in section 3.

## 2. Market-based statistical moments of price and return

The statistical properties of a random variable can be described equally by probability measure, and a set of the *n-th* statistical; moments (Shephard, 1991; Shiryaev, 1999Shreve, 2004). The finite number *m* of the n-th statistical moments for *n=1,2,.m* describes the *m-th* approximation of probability measure. To describe the random properties of price and return we consider their averages and volatilities and describe their dependence on 1st and 2nd statistical moments and correlations of random market trade values and volumes. That dependence emphasizes the impact of market trade randomness on the statistical properties of price and return and explains the restrictions for predictions of price and return probabilities with accuracy that would exceed Gaussian distributions. This section follows Olkhov (2021-2023) and we refer there for further details.

### *2.1 Market-based statistical moments of price*

We assume that market trades with stocks are made at a time $t_i$ with a constant interval $\varepsilon$ between trades:

$$\varepsilon - const \quad ; \quad t_i = t_0 + i\varepsilon \quad ; \quad i = 0,1,\dots \tag{2.1}$$

The interval $\varepsilon$ between trades introduces the initial market time axis division multiple of $\varepsilon$ (2.1). As initial, we consider the time series of trade values $C(t_i)$ and volumes $U(t_i)$ with stocks at times $t_i$ that determine a primitive equation (2.2) for trade price $p(t_i)$:

$$C(t_i) = p(t_i)U(t_i) \quad ; \quad i = 0,1,\dots \tag{2.2}$$

Equation (2.2) defines the market trade price $p(t_i)$ of the stocks of an individual company at $t_i$. The initial time axis division $\varepsilon$ can be equal to a second or even a fraction of a second. The time series of the trade value $C(t_i)$, volume $U(t_i)$ and price $p(t_i)$ are very irregular and of little help for predictions of the stock price at a time horizon $T$ that can be equal to weeks, months, or years. One can consider market time series as random variables during any reasonable time interval $\Delta>>\varepsilon$. For simplicity, we take $\Delta$ as a multiple of $\varepsilon$ (2.3) with $N$ terms of the time series $t_i$ inside $\Delta$. To develop a pricing model at the horizon $T>>\varepsilon$ one should average the initial random market time series over the interval $\Delta$ (2.3):

$$\Delta = N\varepsilon \quad ; \quad N \gg 1 \; ; \quad \varepsilon \ll \Delta < T \tag{2.3}$$

After averaging the initial market time series over $\Delta$ (2.3), one gets more smooth data that can be more useful for forecasting at horizon $T$. Averaged time series introduce a transition from the initial market time axis division that is multiple of $\varepsilon$ to a new one, a rougher time axis division multiple of $\Delta$. Market trades with stocks of any company determine three initial time series of the financial variables that should be taken into account by any pricing model: the trade value $C(t_i)$, volume $U(t_i)$ and price $p(t_i)$ (2.2). The independent definitions for the probabilities of the trade value $C(t_i)$, volume $U(t_i)$, and price $p(t_i)$ that match the equation (2.2) are impossible. We consider the random time series of the trade values $C(t_i)$ and volumes $U(t_i)$ as the primary, which completely determines the stochasticity of the market price $p(t_i)$. To support this statement, we refer to Fox et al. (2017), which provide the perfect methodology for estimating national accounts by the aggregation of additive economic variables as the basis for the definition of non-additive variables such as price, inflation, bank rates, etc. We follow Fox et al. (2017) and consider the additive random variables determined by the time series of trade values $C(t_i)$ and volumes $U(t_i)$ as the basis for describing the random properties of the stock price and return.

Assume that the averaging interval $\Delta$ defines the time axis division $t_k$, $k=0,1,..$ multiple of $\Delta$:

$$\Delta_k = \left[t_k - \frac{\Delta}{2}; t_k + \frac{\Delta}{2}\right] \quad ; \quad t_k = t_0 + \Delta \cdot k \quad ; \quad k = 0, 1, 2, \ldots \tag{2.4}$$

For convenience, we renumber the initial trade time series $t_i$ (2.1; 2.3) and note them as $t_{ik}$, which belong to interval $\Delta_k$ (2.4):

$$t_k - \frac{\Delta}{2} \le t_{ik} \le t_k + \frac{\Delta}{2} \quad ; \quad t_{i+1,k} - t_{ik} = \varepsilon \quad ; \quad t_{i,k+1} - t_{ik} = \Delta \quad ; \quad i = 1, .. N \tag{2.5}$$

Thus, we consider $N$ terms of the time series $t_{ik}$ in each interval $\Delta_k$ (2.4). That allows an equal estimate of the statistical moments of the market trade value $C(t_k;n)$ and volume $U(t_k;n)$ in each averaging interval $\Delta_k$ as (2.6):

$$C(t_k; n) = E[C^n(t_{ik})] \sim \frac{1}{N}\sum_{i=1}^{N} C^n(t_{ik}) \quad ; \quad U(t_k; n) = E[U^n(t_{ik})] \sim \frac{1}{N}\sum_{i=1}^{N} U^n(t_{ik}) \tag{2.6}$$

We use the symbol ~ to highlight that (2.6) defines only assessments of mathematical expectation *E[..]* by a finite number *N* of terms of the time series that belong to the interval $\Delta_k$ (2.5). The *n*-th degree of equation (2.2) at time $t_{ik}$ gives:

$$C^n(t_{ik}) = p^n(t_{ik})\, U^n(t_{ik}) \quad ; \quad n = 1,2, \ldots \tag{2.7}$$

The equations (2.7) help define market-based *n-th* statistical moments of price and return (Olkhov, 2021a; 2022; 2023a). For simplicity in this paper we consider only averages and volatilities of price and return. The derivation of the first four statistical moments of price is given in (Olkhov, 2023a). As market-based average price $a(t_k;1)$ we take volume weighted average price (VWAP) (Berkowitz et al., 1983; Duffie and Dworczak, 2018):

$$a(t_k;1) = E_m[p(t_{ik})] = \frac{1}{\sum_{i=1}^{N} U(t_{ik})} \sum_{i=1}^{N} p(t_{ik})U(t_{ik}) = \frac{\sum_{i=1}^{N} C(t_{ik})}{\sum_{i=1}^{N} U(t_{ik})} = \frac{C(t_k;1)}{U(t_k;1)} \tag{2.8}$$

In (2.8) we denote market-based mathematical expectation as $E_m[..]$ to differ it from (2.6). We define the market-based price volatility $\sigma^2(t_k)$ (2.9) and 2-d statistical moment $a(t_k;2)$ (2.10) of price and refer (Olkhov, 2021a; 2022; 2023a) for details:

$$\sigma^2(t_k) = \frac{\Omega_C^2(t_k) + a^2(t_k;1)\Omega_U^2(t_k) - 2a(t_k;1)corr\{C(t_k)U(t_k)\}}{U(t_k;2)} \tag{2.9}$$

$$a(t_k;2) = \frac{C(t_k;2) + 2a^2(t_k;1)\Omega_U^2(t_k) - 2a(t_k;1)corr\{C(t_k)U(t_k)\}}{U(t_k;2)} \tag{2.10}$$

In (2.9; 2.10) $\Omega_C^2(t_k)$ and $\Omega_U^2(t_k)$ (2.11) denote trade value and trade volume volatilities respectively.

$$\Omega_C^2(t_k) = C(t_k;2) - C^2(t_k;1) \quad ; \quad \Omega_U^2(t_k) = U(t_k;2) - U^2(t_k;1) \tag{2.11}$$

The correlation $corr\{C(t_k)U(t_k)\}$ (2.12) between the trade value $C(t_k)$ and volume $U(t_k)$ during interval $\Delta_k$ (2.4) depends on the joint average $CU(t_k;1)$ (2.13) of the product of the trade value $C(t_k)$ and volume $U(t_k)$:

$$corr\{C(t_k)U(t_k)\} = CU(t_k;1) - C(t_k;1)U(t_k;1) \tag{2.12}$$

$$CU(t_k;1) = E[C(t_{ik})U(t_{ik})] = \frac{1}{N}\sum_{i=1}^{N} C(t_{ik})U(t_{ik}) \tag{2.13}$$

### *2.2 Market-based statistical moments of return*

In this paper we describe the average and volatility of stock return with a time shift $\xi$. We assume that all prices are adjusted to current time $t_0$ and consider the trade equation (2.2) during interval $\Delta_k$ (2.4) as follows:

$$C(t_{ik}) = p(t_{ik})U(t_{ik}) = \frac{p(t_{ik})}{p(t_{ik}-\xi)}\, p(t_{ik}-\xi)U(t_{ik}) = r(t_{ik},\xi)S(t_{ik},\xi) \tag{2.14}$$

We denote return $r(t_{ik},\xi)$ (2.14) as the ratio of price $p(t_{ik})$ at $t_{ik}$ (2.5) to price $p(t_{ik}-\xi)$ in the past at $t_{i,k}-\xi$. For convenience, we take the time shift $\xi$ as a multiple of $\varepsilon$:

$$S(t_{ik},\xi) \equiv p(t_{ik}-\xi)U(t_{ik}) \quad ; \quad r(t_{ik},\xi) \equiv \frac{p(t_{ik})}{p(t_{ik}-\xi)} \quad ; \quad \xi = \varepsilon\, j \tag{2.15}$$

We call the return $r(t_{ik},\xi)$ (2.14; 2.15) an "anticipated" return because it is estimated through the market price time series $p(t_{ik})$ and $p(t_{ik}-\xi)$, but not through real purchases in the past and current sales of stock by a particular investor. We present the market-based average and volatility of the "anticipated" return according to Olkhov (2023a). The description of averages and volatilities of the "actual" return of investors that are estimated as a ratio of the current sale price to the price of the purchases in the past is presented in Olkhov (2023b). To simplify notations, we use the time shift $\xi$ without index $j$ (2.15). We denote $S(t_{ik},\xi)$ as the past value of the volume $U(t_{ik})$ of stocks at a price $p(t_{ik}-\xi)$. Using (2.2; 2.15), we present (2.14) as the return trade equation (2.16):

$$C(t_{ik}) = r(t_{ik},\xi)S(t_{ik},\xi) \tag{2.16}$$

Similar to (2.6), we define the *n-th* statistical moments $S(t_k,\xi;n)$ of the past value $S(t_{ik},\xi)$ (2.15) determined by the volume $U(t_{ik})$ of stocks at the price $p(t_{ik}-\xi)$:

$$S(t_k,\xi;n) = E[S^n(t_{ik},\xi)] \sim \frac{1}{N}\sum_{i=1}^{N} S^n(t_{ik},\xi) = \frac{1}{N}\sum_{i=1}^{N} p^n(t_{ik}-\xi)U^n(t_{ik}) \tag{2.17}$$

The market-based average return $h(t_k;1)$ (2.18), which has the same economic notion as VWAP (2.8), was introduced by Markowitz (1952) 35 years before (Berkowitz et al., 1983). Markowitz defined the portfolio return as "weighted with weights equal to the relative amount invested in security." That definition almost completely reproduces the definition of VWAP but replaces the volume with the past value of the stock. One can consider $N$ market trades during interval $\Delta_k$ (2.4) at time $t_{ik}$ as "securities" with current value $C(t_{ik})$ and past value $S(t_{ik},\xi)$ (2.15). The market-based average return $h(t_k,\xi;1)$ (2.18 – 2.20) takes the form (Olkhov, 2023a):

$$h(t_k,\xi;1) = E_m[r(t_{ik},\xi)] = \frac{1}{\sum_{i=1}^{N} S(t_{ik},\xi)} \sum_{i=1}^{N} r(t_{ik},\xi)S(t_{ik},\xi) \tag{2.18}$$

$$h(t_k,\xi;1) = \frac{\sum_{i=1}^{N} C(t_{ik})}{\sum_{i=1}^{N} S(t_{ik},\xi)} = \frac{C(t_k;1)}{S(t_k,\xi;1)} \tag{2.19}$$

$$C(t_k;1) = h(t_k,\xi;1)S(t_k,\xi;1) \tag{2.20}$$

Similar to (2.9-2.13), the market-based volatility $v^2(t_k,\xi)$ (2.21) of return and the 2-d statistical moment $h(t_k,\xi;2)$ of return take the form (Olkhov, 2023a):

$$v^2(t,\xi) = E_m\left[\left(r(t_{ik},\xi) - h(t_k,\xi;1)\right)^2\right] = h(t_k,\xi;2) - h^2(t_k,\xi;1) \geq 0$$

$$v^2(t_k,\xi) = \frac{\Omega_C^2(t_k) + h^2(t_k,\xi;1)\Omega_S^2(t_k,\xi) - 2h(t_k,\xi;1)corr\{C(t_k)S(t_k,\xi)\}}{S(t_k,\xi;2)} \tag{2.21}$$

$$h(t_k,\xi;2) = \frac{C(t_k;2) + 2h^2(t_k,\xi;1)\Omega_S^2(t_k,\xi) - 2h(t_k,\xi;1)corr\{C(t_k)S(t_k,\xi)\}}{S(t_k,\xi;2)} \tag{2.22}$$

In (2.21; 2.22), we denote the volatility $\Omega_C^2(t_k)$ (2.11) of the current value $C(t_{ik})$ and the volatility $\Omega_S^2(t_k,\xi)$ (2.23) of the past value $S(t_{ik},\xi)$ (2.15) that takes the form:

$$\Omega_S^2(t_k,\xi) = E[(S(t_{ik},\xi) - S(t_k,\xi;1))^2] = S(t_k,\xi;2) - S^2(t_k,\xi;1) \tag{2.23}$$

The correlation $corr\{C(t_k)S(t_k,\xi)\}$ (2.24) between the current value $C(t_k)$ and the past value $S(t_k,\xi)$ (2.15) during the interval $\Delta_k$ (2.4) depends on the joint average $CS(t_k,\xi;1)$ (2.25) of the product of the current value $C(t_k)$ and the past value $S(t_k,\xi)$:

$$corr\{C(t_k)S(t_k,\xi)\} = CS(t_k,\xi;1) - C(t_k;1)S(t_k,\xi;1) \tag{2.24}$$

$$CS(t_k,\xi;1) = E[C(t_{ik})S(t_{ik},\xi)] = \frac{1}{N}\sum_{i=1}^{N} C(t_{ik})S(t_{ik},\xi) \tag{2.25}$$

The first two market-based statistical moments $a(t_k;1)$ (2.8) and $a(t_k;2)$ (2.10) of price and two market-based statistical moments $h(t_k,\xi;1)$ (2.18-2.20) and $h(t_k,\xi;2)$ (2.22) of return describe the Gaussian approximations of the price and return probabilities. The derivation of the first four market-based statistical moments of price and return as functions of statistical moments and correlations of market trade value, volume, and past value is given in (Olkhov, 2023a). However, the complexity of economic relations severely restricts any predictions of price and return probabilities, in the best case, by the Gaussian-type distributions. Thus, the description of higher statistical moments of price and return that can model current probabilities with more accuracy doesn't help for predictions of probabilities.

The predictions of the market-based statistical moments of price and return of stock require knowledge of the market "environment": the estimates of the price and return of other similar stocks traded on the market. To describe the statistical moments of numerous stocks traded on the NYSE or Nasdaq, one should distribute stocks by some parameters to distinguish them from each other. As a parameter that helps distribute different stocks, we select the risk ratings of their issuer companies. In the next section, we explain how the assessments of the risk ratings of issuer companies introduce the notion of risk coordinates in the economic domain and describe the market-based statistical moments of the stock return as functions of risk.

## 3 Statistical moments of the collective price and return as functions of risk

In this section, we describe the dependence of the market-based statistical moments of stock price and return on risk ratings of the economic agents that make the trades. We consider agents' risk ratings as their coordinates in the economic domain (Olkhov, 2016-2020). The major risk agencies, such as Fitch, Moody's, and S&P assess the risk ratings of the majority of stocks, banks, and corporations (Metz and Cantor, 2007; Chane-Kon et al., 2010; Kraemer and Vazza, 2012). Risk agencies use the letter notations *AAA, AA, BB,* and *C* to designate the risk rate. Each rating agency has its own letter grade system to protect and promote their business. However, more than 80 years ago, Durand (1941) proposed the use of numerical risk grades. Indeed, risk ratings are conditional terms that are used as helpful tools for management, investment, and economic modeling. There is no difference in how one denotes a particular risk rating: as a letter *A* or as a number *3.* Primarily, the risk metrics should help describe economic problems but not serve the promotion of a particular business. The use of numeric risk grades can result in a unified methodology for risk assessments by different agencies and can open up wide opportunities for economic and financial modeling. We take Durand's (1941) idea of numeric risk grades and complement it with introducing continuous numeric risk grades. The notions of the most secure and the most risky grades are completely arbitrary, and the symbol *AAA* can easily be replaced by a numeric value. We take the most secure risk grade to be equal to 0 and the most risky grade to be equal to 1. Altogether, we replace the letter-based risk grade symbols *AAA, BB,* and *CC* by continuous numeric risk grades that fill the unit interval [0,1], which we call the economic domain. If one considers the economic system under the action of *J* risks, then the numeric values of agents' ratings fill the economic domain as a unit cube $[0,1]^J$. The description of agents by their risk coordinates in the economic domain gives great advantages for economic and financial modeling and reveals hidden and missed economic factors and processes that impact economic evolution. For simplicity, we consider market trades under the action of a single risk. For definiteness, one can consider the credit ratings of economic agents. We don't discuss here a particular methodology for the assessment of numeric continuous credit risk ratings and consider it a worthy task for the risk rating agencies. Actually, the substitution of the conventional letter designations of risk ratings by numeric continuous risk ratings is like opening Pandora's box of hidden economic complexity. Indeed, market trade performance completely determines economic development. Each trade on the stock market can be described by the buyer, seller, and the traded stock that is issued by a particular company. To describe a single trade, one should specify the risk ratings of at least three economic agents: the buyer, the seller, and the issuer of the stock. Thus, to define ratings of the single trade, we introduce the risk vector $\boldsymbol{x}=(x_1,x_2,x_3)$, which takes values in the economic domain – the unit cube (3.1):

$$\boldsymbol{x} \in [0,1]^3 \;;\quad \boldsymbol{x} = (x_1, x_2, x_3) \quad ;\quad 0 \le x_i \le 1 \;;\; i = 1,2,3 \tag{3.1}$$

As we mentioned in Section 2, we consider the "anticipated" return. An investor can be a seller, a buyer, or even an issuer, depending on his market trade decisions. For certainty, we consider an investor as a buyer and model the "anticipated" return of the current purchases at time $t_{ik}$ with respect to the stock price in the past at time $t_{ik}$-$\xi$. We assume that the components of the risk vector describe ratings: $x_1$ – of the buyer, $x_2$ – of the issuer of stock, and $x_3$ – of the seller. We denote a particular trade of an investor as a buyer by risk coordinates $x_{b1}$, risk coordinates $x_{q2}$ of the stock of a particular issuer $q$, and risk coordinates $x_{s3}$ of a seller. To consider the statistical moments of trade made by the investor as a buyer we sum trades with stock $x_{q2}$ over all sellers with risk coordinates $x_{s3}$ and consider the values $C(t_{ik},\boldsymbol{x}_q)$ and volumes $U(t_{ik},\boldsymbol{x}_q)$ of market trades of a buyer $x_{b1}$ during $\Delta_k$ (2.4) as functions of the risk vector $\boldsymbol{x_q}=(x_{b1},x_{q2})$ in 2-dimensinal economic domain (3.2):

$$\boldsymbol{x} \in [0,1]^2 \;;\quad \boldsymbol{x} = (x_1, x_2) \quad ; \quad 0 \leq x_i \leq 1 \;;\; i = 1,2 \tag{3.2}$$

As we show in Section 2, the trade values $C(t_{ik},\boldsymbol{x}_q)$ and volumes $U(t_{ik},\boldsymbol{x}_q)$ during $\Delta_k$ (2.4) define statistical moments of trade value $C(t_k,\boldsymbol{x}_q;n)$, volume $U(t_k,\boldsymbol{x}_q;n)$ (2.6), average price $a(t_k,\boldsymbol{x}_q;1)$ (2.8), and price volatility $\sigma^2(t_k,\boldsymbol{x}_q)$ (2.9) as functions of risk vector $\boldsymbol{x_q}$. Statistical moments of past value $S(t_k,\xi,\boldsymbol{x}_q;n)$ (2.17), average return $h(t_k,\xi,\boldsymbol{x}_q;1)$ (2.18 – 2.20), and volatility $v^2(t_k,\xi,\boldsymbol{x}_q)$ (2.21) also depend on risk vector $\boldsymbol{x_q}$. As we mentioned above, to predict the averages and volatilities of the price and return of stock issued by a company $q$, one should model the evolution of the market trade of similar stocks. To define the market environment, let us collect the values and volumes of the trades made by all investors with stocks in the neighborhood $dV(\boldsymbol{x})$ (3.3) of point $\boldsymbol{x}=(x_1,x_2)$ of the economic domain (3.2). In 2022, the NYSE traded around 2500 stocks, and the Nasdaq traded almost 3600 stocks of domestic and international companies (Statista, 2023). We consider that the number $Q$ of companies in the market is high $Q>>1$, and denote $Q(\boldsymbol{x})$ as the number of stocks with risk coordinates near point $\boldsymbol{x}$ (3.3). Let us choose a scale $d<1$ that defines a small space $dV(\boldsymbol{x})$ in the economic domain (3.2):

$$0 < d < 1 \;;\; dV(\boldsymbol{x}) \sim d^2 \;;\; \boldsymbol{x}_q \in dV(\boldsymbol{x}) \leftrightarrow x_i - \frac{d}{2} \leq x_{qi} \leq x_i + \frac{d}{2} \;;\; \boldsymbol{x} = (x_1, x_2) \tag{3.3}$$

The choice of the scale $d$ allows at time $t_{ik}$ select the buyers of stocks issued by the companies. We assume that the buyers have risks $x_{b1}$, the issuers of stocks have risks $x_{q2}$, and they define the market deal with the risk coordinates $\boldsymbol{x}_q=(x_{b1},x_{q2})$ inside a $dV(\boldsymbol{x})$ (3.3). To define the first two statistical moments of the *collective* values and volumes of trade inside $dV(\boldsymbol{x})$ (3.3) for $m=1,2$ we sum the $m$-th degree of values $C^m(t_{ik},\boldsymbol{x}_q)$ and volumes $U^m(t_{ik},\boldsymbol{x}_q)$ as:

$$C(t_{ik}, \boldsymbol{x}; m) = \sum_{x_q \in dV(x)} C^m(t_{ik}, \boldsymbol{x}_q) \quad ; \quad U(t_{ik}, \boldsymbol{x}; m) = \sum_{x_q \in dV(x)} U^m(t_{ik}, \boldsymbol{x}_q) \tag{3.4}$$

The value $C(t_{ik},\boldsymbol{x};m)$ (3.4) at a time $t_{ik}$ equals the sum of the $m$-th degrees, $m=1,2$ of values of the stocks of companies $q$ purchased by all investors with coordinates $\boldsymbol{x}_q$ inside a volume $dV(\boldsymbol{x})$ (3.3). The volume $U(t_{ik},\boldsymbol{x};m)$ (3.4) equals the corresponding sum of the $m$-th degrees of all trade volumes at a time $t_{ik}$. Relations (3.4) for $m=1,2$ transfer the description of the trade values $C^m(t_{ik},\boldsymbol{x}_q)$ and volumes $U^m(t_{ik},\boldsymbol{x}_q)$ as functions of coordinates $\boldsymbol{x}_q$ of a buyer $x_{b1}$ of the stocks of a company $x_{q2}$ to the description of the *collective* trade values $C(t_{ik},\boldsymbol{x};m)$ and volumes $U(t_{ik},\boldsymbol{x};m)$ as functions of coordinates $\boldsymbol{x}$. We introduce a similar definition of the *collective* past values $S^m(t_{ik},\xi,\boldsymbol{x}_q;m)$:

$$S(t_{ik},\xi,\boldsymbol{x};m)=\sum_{x_q\in dV(x)} S^m\left(t_{ik},\xi,\boldsymbol{x}_q\right) \quad ; \quad m=1,2 \tag{3.5}$$

To derive statistical moments of the *collective* values and volumes (3.4; 3.5) of market trade that define the market-based averages and volatilities of price and return as functions of time and risk coordinates $\boldsymbol{x}$, one should average (3.4; 3.5) over the *collective* time averaging interval $\Delta_{\boldsymbol{x}}$. The choice of the *collective* averaging interval $\Delta_{\boldsymbol{x}}$ is not a simple problem. The interval $\Delta$ (2.3-2.5) determines the averaging of the trade values and volumes of the stocks of a particular company $q$. The *collective* averaging interval $\Delta_{\boldsymbol{x}}$ should be $\Delta_{\boldsymbol{x}} \geq \Delta$, and we assume that $\Delta_{\boldsymbol{x}}$ is the same for all points $\boldsymbol{x}$ in the economic domain. The choice of the averaging interval $\Delta_{\boldsymbol{x}}$ introduces a new time axis division $\tau_k$, which describes the statistical moments of the *collective* purchases of all investors and all stocks with risk coordinates inside $dV(\boldsymbol{x})$ (3.3) and averaged over $\Delta_{\boldsymbol{x}}$. For simplicity, we take $\Delta_x$ as a multiple of $\Delta$ (2.3):

$$\Delta_x = k_x\Delta = k_x N\varepsilon \quad ; \quad k_x=1,2,.. \quad ; \quad \Delta = N\varepsilon \tag{3.6}$$

$$\tau_k = t_0 + k\,\Delta_x \quad ; \quad k=1,2,.. \quad ; \quad \tau_k - \frac{\Delta_x}{2} \leq t_{ik} \leq \tau_k + \frac{\Delta_x}{2} \quad ; \quad i=0,1,\dots k_x N \tag{3.7}$$

Similar to (2.4; 2.5), we renumber the initial time series $t_i$ (2.1) so that each interval $\Delta_{\boldsymbol{x}}$ (3.6; 3.7) contains the same number $k_xN$ of terms of the trades (3.4; 3.5). The choice of the interval $\Delta_{\boldsymbol{x}}$ helps to average the *collective* values $C(t_{ik},\boldsymbol{x};m)$, volumes $U(t_{ik},\boldsymbol{x};m)$ (3.4), and the *collective* past values $S(t_{ik},\xi,\boldsymbol{x};m)$ (3.5). We determine the $m$-th statistical moments of the *collective* trade values $C(t_{ik},\boldsymbol{x};m)$ at time $\tau_k$ averaged over $\Delta_{\boldsymbol{x}}$ as:

$$C(\tau_k,\boldsymbol{x};m)=\frac{1}{k_xN}\sum_{i=1}^{k_xN} C(t_{ik},\boldsymbol{x};m)=\frac{1}{k_xN}\sum_{i=1}^{k_xN}\sum_{x_q\in dV(x)} C^m\left(t_{ik},\boldsymbol{x}_q\right) \tag{3.8}$$

If one changes the order of sums in (3.8) then:

$$C(\tau_k,\boldsymbol{x};m)=\sum_{x_q\in dV(x)}\frac{1}{k_xN}\sum_{i=1}^{k_xN} C^m(t_{ik},\boldsymbol{x})=\sum_{x_q\in dV(x)} C\left(\tau_k,\boldsymbol{x}_q;m\right) \quad ; \quad m=1,2 \tag{3.9}$$

$$C\left(\tau_k,\boldsymbol{x}_q;m\right)=\frac{1}{k_xN}\sum_{i=1}^{k_xN} C^m\left(t_{ik},\boldsymbol{x}_q\right) \tag{3.10}$$

The relations (3.10) at time $\tau_k$ (3.7) denote the $m$-th statistical moments $C(\tau_k,\boldsymbol{x}_q;m)$ of the values of the stocks of company $x_{q2}$ of the purchases by the investor $x_{b1}$ averaged over $\Delta_x$. Thus, the $m$-th statistical moments $C(\tau_k,\boldsymbol{x};m)$ (3.8; 3.9) of the *collective* trade values at point $\boldsymbol{x}$ equals the sum of the *m-th* statistical moments of the trade values of trades of the buyer $x_{b1}$ with stocks of all companies $x_{q2}$ inside the $dV(\boldsymbol{x})$ (3.3) averaged over $\Delta_x$ (3.8; 3.9). The same meaning have the $m$-th statistical moments of the *collective* volumes $U(\tau_k,\boldsymbol{x};m)$ (3.11; 3.12):

$$U(\tau_k,\boldsymbol{x};m)=\frac{1}{k_xN}\sum_{i=1}^{k_xN}U(t_{ik},\boldsymbol{x};m) \tag{3.11}$$

$$U(\tau_k,\boldsymbol{x};m)=\sum_{\boldsymbol{x}_q\in dV(x)}\frac{1}{k_xN}\sum_{i=1}^{k_xN}U^m\left(t_{ik},\boldsymbol{x}_q\right)=\sum_{\boldsymbol{x}_q\in dV(x)}U\left(\tau_k,\boldsymbol{x}_q;m\right) \tag{3.12}$$

Relations (3.8-3.12) define the $m$-th statistical moments of the collective trade values $C(\tau_k,\boldsymbol{x};m)$ and volumes $U(\tau_k,\boldsymbol{x};m)$ as functions of time $\tau_k$ and coordinates $\boldsymbol{x}$ in the $dV(\boldsymbol{x})$ (3.3). Similar relations determine the statistical moments of the collective past values $S(\tau_k,\xi,\boldsymbol{x};m)$:

$$S(\tau_k,\xi,\boldsymbol{x};m)=\frac{1}{k_xN}\sum_{i=1}^{k_xN}S(t_{ik},\xi,\boldsymbol{x};m)=\frac{1}{k_xN}\sum_{i=1}^{k_xN}\sum_{\boldsymbol{x}_q\in dV(x)}S^m\left(t_{ik},\xi,\boldsymbol{x}_q\right) \tag{3.13}$$

$$S(\tau_k,\xi,\boldsymbol{x};m)=\sum_{\boldsymbol{x}_q\in dV(x)}\frac{1}{k_xN}\sum_{i=1}^{k_xN}S^m\left(t_{ik},\xi,\boldsymbol{x}_q\right)=\sum_{\boldsymbol{x}_q\in dV(x)}S\left(\tau_k,\boldsymbol{x}_q;m\right) \tag{3.14}$$

We highlight that (3.8 -3.14) give the approximations of the statistical moments by a finite number $k_xN$ of terms of time series. We use the relations (3.8-3.14), similar to (2.8-2.13) and (2.18-2.26), and determine the market-based averages and volatilities of the *collective* price and return as functions of $(\tau_k,\boldsymbol{x})$. The collective average $a(\tau_k,\boldsymbol{x};1)$ and volatility $\sigma^2(\tau_k,\boldsymbol{x})$ of price of all trades of buyers $x_{b1}$ with stocks of companies $x_{q2}$ inside $dV(\boldsymbol{x})$ (3.3) and averaged over $\Delta_x$ (3.8; 3.9) take the form:

$$a(\tau_k,\boldsymbol{x};1)=\frac{C(\tau_k,\boldsymbol{x};1)}{U(\tau_k,\boldsymbol{x};1)} \tag{3.15}$$

$$\sigma^2(\tau_k,\boldsymbol{x})=\frac{\Omega_C^2(\tau_k,\boldsymbol{x})+a^2(\tau_k,\boldsymbol{x};1)\Omega_U^2(\tau_k,\boldsymbol{x})-2a(\tau_k,\boldsymbol{x};1)corr\{C(\tau_k,\boldsymbol{x})U(\tau_k,\boldsymbol{x})\}}{U(\tau_k,\boldsymbol{x};2)} \tag{3.16}$$

The 2-d market-based price statistical moment $a(\tau_k,\boldsymbol{x};2)$:

$$a(\tau_k,\boldsymbol{x};2)=\frac{C(\tau_k,\boldsymbol{x};2)+2a^2(\tau_k,\boldsymbol{x};1)\Omega_U^2(\tau_k,\boldsymbol{x})-2a(\tau_k,\boldsymbol{x};1)corr\{C(\tau_k,\boldsymbol{x})U(\tau_k,\boldsymbol{x})\}}{U(\tau_k,\boldsymbol{x};2)} \tag{3.17}$$

The volatilities of trade value and volume (3.18) and their correlation (3.19) take the form similar to (2.11-2.13):

$$\Omega_C^2(\tau_k,\boldsymbol{x})=C(\tau_k,\boldsymbol{x};2)-C^2(\tau_k,\boldsymbol{x};1)\quad;\quad\Omega_U^2(\tau_k,\boldsymbol{x})=U(\tau_k,\boldsymbol{x};2)-U^2(\tau_k,\boldsymbol{x};1) \tag{3.18}$$

$$corr\{C(\tau_k,\boldsymbol{x})U(\tau_k,\boldsymbol{x})\}=CU(\tau_k,\boldsymbol{x};1)-C(\tau_k,\boldsymbol{x};1)U(\tau_k,\boldsymbol{x};1) \tag{3.19}$$

The joint average $CU(\tau_k,\boldsymbol{x};1)$ of the trade value and volume as a function of $(\tau_k,\boldsymbol{x})$ takes the form:

$$CU(\tau_k,\boldsymbol{x};1)=\sum_{\boldsymbol{x}_q\in dV(x)}\frac{1}{k_xN}\sum_{i=1}^{k_xN}C\left(t_{ik},\boldsymbol{x}_q\right)U\left(t_{ik},\boldsymbol{x}_q\right) \tag{3.20}$$

The average $h(\tau_k,\xi,\boldsymbol{x};1)$ and volatility $\nu^2(\tau_k,\xi,\boldsymbol{x})$ of the *collective* return as functions of $(\tau_k,\xi,\boldsymbol{x})$ take the form similar to (2.18-2.22) :

$$h(\tau_k,\xi,\boldsymbol{x};1)=\frac{C(\tau_k,\boldsymbol{x};1)}{S(\tau_k,\xi,\boldsymbol{x};1)} \tag{3.21}$$

$$\nu^2(\tau_k,\xi,\boldsymbol{x})=\frac{\Omega_C^2(\tau_k,\boldsymbol{x})+h^2(\tau_k,\xi,\boldsymbol{x};1)\Omega_S^2(\tau_k,\xi,\boldsymbol{x})-2h(\tau_k,\xi,\boldsymbol{x};1)corr\{C(\tau_k,\boldsymbol{x})S(\tau_k,\xi,\boldsymbol{x})\}}{S\left(\tau_k,\xi,\boldsymbol{x};2\right)} \tag{3.22}$$

The 2-d market-based statistical moment $h(\tau_k,\xi,\boldsymbol{x};2)$ of return takes the form:

$$h(\tau_k, \xi, \boldsymbol{x}; 2) = \frac{C(\tau_k, \boldsymbol{x}; 2) + 2h^2(\tau_k, \xi, \boldsymbol{x}; 1)\Omega_S^2(\tau_k, \xi, \boldsymbol{x}) - 2h(\tau_k, \xi, \boldsymbol{x}; 1) corr\{C(\tau_k, \boldsymbol{x})S(\tau_k, \xi, \boldsymbol{x})\}}{S(\tau_k, \xi, \boldsymbol{x}; 2)} \quad (3.23)$$

The volatility of the past trade value $\Omega_S^2(\tau_k,\xi,\boldsymbol{x})$ (3.24) and correlation $corr\{C(\tau_k,\boldsymbol{x})S(\tau_k,\xi,\boldsymbol{x})\}$ (3.25) between current $C(\tau_k,\boldsymbol{x})$ and past $S(\tau_k,\xi,\boldsymbol{x})$ trade values take the form:

$$\Omega_S^2(\tau_k, \xi, \boldsymbol{x}) = S(\tau_k, \xi, \boldsymbol{x}; 2) - S^2(\tau_k, \xi, \boldsymbol{x}; 1) \quad (3.24)$$

$$corr\{C(\tau_k, \boldsymbol{x})S(\tau_k, \xi, \boldsymbol{x})\} = CS(\tau_k, \xi, \boldsymbol{x}; 1) - C(\tau_k, \boldsymbol{x}; 1)S(\tau_k, \xi, \boldsymbol{x}; 1) \quad (3.25)$$

The joint average $CS(\tau_k,\xi,\boldsymbol{x};1)$ of the current and past trade values takes the form:

$$CS(\tau_k, \xi, \boldsymbol{x}; 1) = \sum_{\boldsymbol{x}_q \in dV(\boldsymbol{x})} \frac{1}{k_x N} \sum_{i=1}^{k_x N} C(t_{ik}, \boldsymbol{x}_q) S(t_{ik}, \xi, \boldsymbol{x}_q) \quad (3.26)$$

The averages and volatilities of price and return (3.15-3.26) as functions of risk coordinates $\boldsymbol{x}$ highlight the important relations that impact the assessments and predictions of price and return of stock of a particular company. Indeed, the average $h(\tau_k,\xi,\boldsymbol{x};1)$ (3.21) and volatility $v^2(\tau_k,\xi,\boldsymbol{x})$ (3.22) describe the *collective* return of stocks of companies with risk coordinates $x_{q2}$ purchased by investors with coordinates $x_{b1}$ inside a small neighborhood $dV(\boldsymbol{x})$ (3.3) of a point $\boldsymbol{x}$ of the economic domain (3.2). The comparisons between the average and volatility of return of stock of a selected company and the average and volatility of *collective* return inside a small neighborhood $dV(\boldsymbol{x})$ (3.3) help forecast the future values of average and volatility of return. On the other hand, the above relations illuminate the dependence of the averages and volatilities of price and return of a particular company on the averages and volatilities of the *collective* price and return of stocks of all companies in a small neighborhood $dV(\boldsymbol{x})$ (3.3). The market trades of stock of a particular company determine the price and return of stock of that company, and the *collective* market trades determine the averages and volatilities of the *collective* price and return.

It is obvious that the predictions of *collective* price and return in a small neighborhood $dV(\boldsymbol{x})$ (3.3) as functions of $(\tau_k,\boldsymbol{x})$ require assessments of their "environment" - the *joint* price and return of the market as a whole.

## 4 The statistical moments of the whole market

To describe the market-based statistical moments of the *joint* price and return of the whole market, one should define the trade values, volumes, and past values of the stocks of all companies in the market. We define the *joint m-th* trade values $C(t_{ik};m)$, volumes $U(t_{ik};m)$, and past values $S(t_{ik},\xi;m)$ as sums over all investors and all $Q$ companies traded on the whole market at time $t_{ik}$:

$$C(t_{ik}; m) = \sum_{\boldsymbol{x}_q} C^m(t_{ik}, \boldsymbol{x}_q) \ ; \ U(t_{ik}; m) = \sum_{\boldsymbol{x}_q} U^m(t_{ik}, \boldsymbol{x}_q) \quad (4.1)$$

$$S(t_{ik}, \xi; m) = \sum_{\boldsymbol{x}_q} S^m(t_{ik}, \xi, \boldsymbol{x}_q) \quad ; \quad m = 1,2 \quad (4.2)$$

The sums in (4.1; 4.2) denote sums over risk coordinates $\boldsymbol{x}_q$ of the whole economic domain (3.2), or equally, over all investors and all companies in the market. To smooth variations of the time series (4.1; 4.2), one should choose the time averaging interval $\Delta_m$ that defines a characteristic time of changes in the *joint* trades of the whole market. It is reasonable that the characteristic time $\Delta_m$ of the *joint* market trades should be longer than the characteristic time $\Delta_x$ of changes of the trades in the neighborhood of a point $\boldsymbol{x}$. For simplicity, we take the *joint* averaging interval $\Delta_m$ as:

$$\Delta_m \geq \Delta_x \geq \Delta \quad ; \quad \Delta_m = k_m \Delta_x = k_m k_x \Delta = k_m k_x N \varepsilon \quad ; \quad k_x, k_m = 1,2,.. \tag{4.3}$$

The *joint* averaging interval $\Delta_m$ introduces a time axis division $\mu_k$ multiple of $\Delta_m$:

$$\mu_k = t_0 + k\,\Delta_m \quad ; \quad k = 1,.. \quad ; \quad \mu_k - \frac{\Delta_m}{2} \leq t_{ik} \leq \mu_k + \frac{\Delta_m}{2} \quad ; \quad i = 0,1,\dots k_m k_x N \tag{4.4}$$

Similar to (3.8-3.14) for $m=1,2$ obtain the $m$-th statistical moments of the trade values, volumes, and past values of the *joint* market trades:

$$C(\mu_k; m) = \frac{1}{k_m k_x N} \sum_{i=1}^{k_m k_x N} C(t_{ik}; m) = \frac{1}{k_m k_x N} \sum_{i=1}^{k_m k_x N} \sum_{x_q} C^m(t_{ik}, \boldsymbol{x}_q) = \sum_{x_q} C(\mu_k, \boldsymbol{x}_q; m) \tag{4.5}$$

$$U(\mu_k; m) = \sum_{x_q} \frac{1}{k_m k_x N} \sum_{i=1}^{k_m k_x N} U^m(t_{ik}, \boldsymbol{x}_q) = \sum_{x_q} U(\mu_k, \boldsymbol{x}_q; m) \tag{4.6}$$

$$S(\mu_k, \xi; m) = \frac{1}{k_m k_x N} \sum_{i=1}^{k_m k_x N} S(t_{ik}, \xi; m) = \frac{1}{k_m k_x N} \sum_{i=1}^{k_m k_x N} \sum_{x_q} S^m(t_{ik}, \xi, \boldsymbol{x}_q) \tag{4.7}$$

The relations (4.5-4.7) define the averages and volatilities of the joint price and return of the whole market. The average $a(\mu_k;1)$, the second statistical moment $a(\mu_k;2)$, and the volatility $\sigma^2(\mu_k)$ of the joint price of the whole market take the form similar to (3.15-3.17):

$$a(\mu_k; 1) = \frac{C(\mu_k;1)}{U(\mu_k;1)} \quad ; \quad a(\mu_k; 2) = \frac{C(\mu_k;2) + 2a^2(\mu_k;1)\Omega_U^2(\mu_k) - 2a(\mu_k;1)corr\{C(\mu_k)U(\mu_k)\}}{U(\mu_k;2)} \tag{4.8}$$

$$\sigma^2(\mu_k) = \frac{\Omega_C^2(\mu_k) + a^2(\mu_k;1)\Omega_U^2(\mu_k) - 2a(\mu_k;1)corr\{C(\mu_k)U(\mu_k)\}}{U(\mu_k;2)} \tag{4.9}$$

Volatilities (4.10) and correlations (4.11) of the trade value and volume take the form similar to (3.18-3.20):

$$\Omega_C^2(\mu_k) = C(\mu_k; 2) - C^2(\mu_k; 1) \quad ; \quad \Omega_U^2(\mu_k) = U(\mu_k; 2) - U^2(\mu_k; 1) \tag{4.10}$$

$$corr\{C(\mu_k)U(\mu_k)\} = CU(\mu_k; 1) - C(\mu_k; 1)U(\mu_k; 1) \tag{4.11}$$

The joint average $CU(\mu_k;1)$ of the trade values and volume takes the form:

$$CU(\mu_k; 1) = \sum_{x_q} \frac{1}{k_m k_x N} \sum_{i=1}^{k_m k_x N} C(t_{ik}, \boldsymbol{x}_q) U(t_{ik}, \boldsymbol{x}_q) \tag{4.12}$$

The average $h(\mu_k,\xi;1)$, the second statistical moment $h(\mu_k,\xi;2)$, and volatility $v^2(\mu_k,\xi)$ of the joint return of the whole market take the form similar to (3.21-3.23):

$$h(\mu_k, \xi; 1) = \frac{C(\mu_k;1)}{S(\mu_k,\xi;1)} \tag{4.13}$$

$$h(\mu_k, \xi; 2) = \frac{C(\mu_k;2) + 2h^2(\mu_k,\xi;1)\Omega_S^2(\mu_k,\xi) - 2h(\mu_k,\xi;1)corr\{C(\mu_k)S(\mu_k,\xi)\}}{S(\mu_k,\xi;2)} \tag{4.14}$$

$$\nu^2(\mu_k,\xi) = \frac{\Omega_C^2(\mu_k)+h^2(\mu_k,\xi;1)\Omega_S^2(\mu_k,\xi)-2h(\mu_k,\xi;1)corr\{C(\mu_k)S(\mu_k,\xi)\}}{S(\mu_k\xi;2)} \tag{4.15}$$

Volatility (4.16) of the past trade value and correlations (4.17) of the current and past trade value take the form similar to (3.24-3.26):

$$\Omega_S^2(\mu_k,\xi) = S(\mu_k,\xi;2) - S^2(\mu_k,\xi;1) \tag{4.16}$$

$$corr\{C(\mu_k)S(\mu_k,\xi)\} = CS(\mu_k,\xi;1) - C(\mu_k;1)S(\mu_k,\xi;1) \tag{4.17}$$

The joint average $CS(\mu_k,\xi;1)$ of the current and past trade values takes the form:

$$CS(\mu_k,\xi;1) = \sum_{\boldsymbol{x}_q} \frac{1}{k_m k_x N} \sum_{i=1}^{k_m k_x N} C(t_{ik},\boldsymbol{x}_q) S(t_{ik},\xi,\boldsymbol{x}_q) \tag{4.18}$$

At the end of this section, we highlight the importance of the four consecutive time axis divisions determined by the four time intervals $\varepsilon << \Delta \leq \Delta_x \leq \Delta_m$. The smallest interval $\varepsilon$ is determined by the frequency of market trading. It introduces the initial market trade time series at $t_i$ (2.1). The scale $\Delta$ determines the time averaging interval for the assessments of statistical moments of market trade and return of a particular investor with stock of a particular company and introduces a new time axis division $t_k$ (2.4; 2.5) a multiply of $\Delta$. For simplicity, we assume that $\Delta$ is the same for all stocks traded on the market and that $\Delta=N\varepsilon$ (2.3). The sum of trades of all investors with the stocks of all companies with risk coordinates $\boldsymbol{x}_q$ in the neighborhood of point $\boldsymbol{x}$ of the economic domain transfers the description of the statistical moments of market trade, price, and return of stocks of a particular investor and particular company to the description of the statistical moments of *collective* trade, price, and return as functions of coordinates $\boldsymbol{x}$. The *collective* trade value, volume, price, and return of stocks with risk coordinates in the neighborhood of point $\boldsymbol{x}$ change more slowly than the trade value, volume, and return of the individual stocks purchased by the individual investor. Hence, the effective averaging of the time series of the *collective* trade value, volume, and return near point $\boldsymbol{x}$ can require a time interval $\Delta_x$ that is longer than the interval $\Delta$. For convenience, we take the time scale $\Delta_x$ as $\Delta_x=k_x\,\Delta=k_x\,N\varepsilon$, and $\Delta_x$ introduces a new time axis division $\tau_k$ (3.7) as a multiple of $\Delta_x$. The time series $\tau_k$ (3.7) describes the statistical moments of trade value, volume, and return averaged over $\Delta_x$. Finally, the collective trade and return of the whole stock market determine the time averaging interval $\Delta_m$. The change in the trade of the whole market is slower than the change in the *collective* trade near point $\boldsymbol{x}$. Thus, the market interval $\Delta_m$ should be longer than $\Delta_x$. We take the interval $\Delta_m$ (4.3), which determines the time averaging of the *joint* trades of all investors and of stocks of all companies on the whole market, as $\Delta_m=k_m\,\Delta_x = k_m\,k_x\,\Delta=k_m k_x N\varepsilon$ (4.3). The market interval $\Delta_m$ introduces the market time axis division $\mu_k$ (4.4), which determines the time series of the statistical moments of trade value, volume, price, and return of all investors and stocks traded at the whole market. These time series describe the financial problems of the stock market with different accuracy. The different choices of the averaging time intervals result in different approximations of financial markets.

## 5 The complexity of predictions of statistical moments of price and return

The continuous economic media approximation describes the transition from modeling the economic variables of individual agents to the description of the *collective* variables as functions of risk coordinates $\boldsymbol{x}$ (Olkhov, 2016 – 2020). As the *collective* variables, one can consider the statistical moments of trade value $C(\tau_k,\boldsymbol{x};m)$ (3.8), volume $U(\tau_k,\boldsymbol{x};m)$ (3.12), and past value $S(\tau_k,\xi,\boldsymbol{x};m)$ (3.14) determined as sums of the *m-th* degree of trades of stocks with coordinates inside $dV(\boldsymbol{x})$ (3.3) and averaged over $\Delta_x$ (3.6; 3.7). One can denote the *m-th* statistical moments of trade as the average *collective* trade variables of the *m-th* order. For $m=1,2$, the *collective* variables $C(\tau_k,\boldsymbol{x};m)$ (3.8), $U(\tau_k,\boldsymbol{x};m)$ (3.12), and $S(\tau_k,\xi,\boldsymbol{x};m)$ (3.14) define the averages and volatilities of the price and return (3.15-3.26). In the case of the whole stock market, the relations (4.5-4.7) define the *joint* trade value, volume, and past value.

The change of agents' risk ratings due to economic, financial, and other factors causes a change of their risk coordinates and results in the motion of agents in the economic domain. Each agent carries its own set of economic variables. The c*ollective* motion of agents in the economic domain generates the flows of agents' *collective* economic variables. The equations that are somewhat alike to the equations of flows of fluids (App.) (Olkhov, 2018-2020) describe the evolution of the *collective* variables (3.8; 3.12; 3.14). However, the nature and laws of economic flows have nothing in common with physical hydrodynamics, and we believe any direct parallels between them make no sense.

An investor who evaluates the forecasts at horizon $T$ of the probabilities of price and return of stocks of a particular company $q$ should follow the path we described above, but in reverse order. The investor should start with the forecast at horizon $T$ of the future dynamics, statistical moments, volatilities, and correlations of the *joint* market trade, price, and return of the whole market. The time axis division of this forecast is a multiple of $\Delta_m$. One should consider this forecast as a slow-changing economic environment and use it for the description of the change of the *collective* variables (4.10; 4.13; 4.14), the description of the statistical moments, the volatilities, and correlations of the *collective* trade, price, and return as functions of risk coordinate $\boldsymbol{x}$ in the economic domain (4.1). This forecast should be evaluated for the time axis division multiple of $\Delta_x \leq \Delta_m$. After having these two forecasts, the investor could try to predict at horizon $T$ the averages and volatilities of the price and return of stocks of a particular company $q$ with risk coordinates $\boldsymbol{x}_q$ in the neighborhood of point $\boldsymbol{x}$ of the economic domain.

Any amount of economic, financial, or market data "today" can help assess only approximations of current probabilities of price and return. The doubtfulness of these approximations determines the uncertainty in forecasts and results in future financial losses. To increase the accuracy of the forecasts of price and return probabilities, investors should predict more statistical moments. However, predictions of the 3-d and 4-th statistical moments of price and return depend on predictions of the corresponding 3-d and 4-th statistical moments of market trade. That, in turn, depends on the modeling of macroeconomic variables of the 3-d and 4-th orders at horizon $T$.

On that path, investors will face irresistible economic obstacles that limit the accuracy of any forecasts of market-based probabilities of price and return. We highlight only two challenges, among many others. The first is the lack of economic theory of the second order, which describes the evolution of macroeconomic variables composed of sums of trade values and volumes of the *1-st* and *2-d* degrees. One should create a methodology for collecting econometric data similar to Fox et al. (2017) and establish a second-order economic theory that will be twice as complex and general as the current one. The second obstacle concerns the complexity and uncertainty of the assessments of the risk rating coordinates of economic agents. The current difficulties of agents' risk estimates that are based on the proceeding of agents' variables of the 1-st order would be doubled if one should take into account agents' variables of the 1-st and the 2-d orders. The indeterminacy of agents' risk rating assessments would be projected into a higher inaccuracy of forecasts of the *collective* trade statistical moments as functions of risk coordinates $\boldsymbol{x}$ and predictions of the *joint* trade statistical moments of the whole market. That in turn will increase the uncertainty of predictions of averages and volatilities of price and return.

We believe that in the coming decades, the capacity for predictions of market-based price and return statistical moments will be bounded, in the best case, by the first two, and thus any predictions of price and return probabilities will be no more than Gaussian.

## 6 Conclusion

This paper brings to the table the economic obstacles that limit the accuracy of any forecasts of the market-based probabilities of price and return of stocks of a company by a Gaussians' distributions. In its turn, the Gaussians' forecasts of the price and return probabilities are the core issues that determine the credibility of all asset pricing models and portfolio theories.

Investors should keep in mind that the number of predicted statistical moments of market trade is the main factor that limits the accuracy of the forecasts of price and return probabilities. One can ignore the complexity of forecasting the statistical moments but cannot overcome or solve the problem. The predictions of the probability of return that ignore the dependence of the m-th statistical moments on the economic theory of the *m*-th order will have such high uncertainty and doubtfulness that they can be harmful for investors. Currently, the lack of research on the *second-order* economic theory limits the accuracy of predictions of price and return probabilities, in the best case, by Gaussian approximations. Even the predictions of Gaussian probability require forecasts of the 2-d statistical moments and correlations of market trade values and volumes. Each step beyond Gaussian probabilities needs a lot of econometric and theoretical studies.

We emphasize that the above rather complex model doesn't take into account a lot of extra factors that significantly impact and complicate the description of market trades, statistical moments, and price probability. In particular, one should take into account the dependence of market trades on the *expectations* of the sellers and buyers of stocks and that will increase the complexity of the model by many times. The first approximations of the impact of the *collective* expectations on market trade are presented in Olkhov (2019).

We believe that a general look at the problem of the accuracy of predictions of price and return probabilities will generate research interest and further studies. However, the exact future of the market-based probabilities of stock price and return is reliably hidden from investors and researchers by the complexities of economic reality.

## Appendix

### Equations of motion in the economic domain

In this Appendix, we briefly consider the equations of motion that describe the *collective* trade values $C(\tau_k,\boldsymbol{x};m)$ (3.8) in the neighborhood of point $\boldsymbol{x}$ in the economic domain (3.2). For simplicity, instead of discrete time $\tau_k$, we consider the approximation of continuous time $t$. The derivation of the equations of motion of the continuous economic media approximation has parallels to the derivation of the conventional equations of continuous mechanics (Childress, 2009). We consider $C(t,\boldsymbol{x};m)$ as a function of coordinate $\boldsymbol{x}$ in the economic domain (3.2). To derive the equations on $C(t,\boldsymbol{x};m)$ as a function of $t$ and $\boldsymbol{x}$, we consider their possible change with time $t$. To explain the origin of such a change, we refer to the risk transition matrices that are estimated for the largest banks and corporations by the major risk-rating agencies (Metz and Cantor, 2007; Moody's, 2009; Fitch, 2017; S&P, 2018). The risk transition matrices determine the probabilities $a_{ij}$ that an agent' risk during a time interval $\Theta$ can change from rating $\boldsymbol{x}_i$ to $\boldsymbol{x}_j$. As agents here, we consider the investor with risk $x_{b1}$ and the issuer of stocks with risk coordinates $x_{q2}$. We use the risk coordinates of the investor and the issuer to determine the risk coordinates of the particular trade with the risk vector $\boldsymbol{x}_q=(x_{b1},x_{q2})$. The changes of risk coordinates of the investor and the issuer result in changes of risk coordinates of the trades.

If one replaces the usual letter designations of the risk ratings with the proposed numeric ones, then the transition matrices can determine the motion in the economic domain with a particular velocity (Olkhov, 2016-2020). Indeed, the transition time $\Theta$ for numeric continuous rating $\boldsymbol{x}_i$ to $\boldsymbol{x}_j$ defines the interval $\boldsymbol{l}_{ij}$ and the velocity $\boldsymbol{v}_{ij}$ between $x_i$ and $\boldsymbol{x}_j$:

$$\boldsymbol{l}_{ij} = \boldsymbol{x}_j - \boldsymbol{x}_i \quad ; \quad \boldsymbol{v}_{ij} = \frac{\boldsymbol{l}_{ij}}{\Theta} \tag{A.1}$$

Taking probabilities $a_{ij}$ of the transition from $\boldsymbol{x}_i$ to $\boldsymbol{x}_j$ during $\Theta$ with the velocity $\boldsymbol{v}_{ij}$ (A.1) one assesses the mean velocity $\boldsymbol{v}(t,\boldsymbol{x}_i)$ of agent at point $\boldsymbol{x}_i$:

$$\boldsymbol{v}(t,\boldsymbol{x}_i) = \sum_{j=1}^{K} \boldsymbol{v}_{ij} a_{ij} = \frac{1}{\Theta}\sum_{j=1}^{K} l_{ij} a_{ij} \quad ; \quad \sum_{j=1}^{K} a_{ij} = 1 \tag{A.2}$$

In (A.2), $K$ denotes the number of the different risk grades of the transition matrix $K$x$K$. Now, we assume that the trade at point $\boldsymbol{x}_q$, which is determined by the investor's risk $x_{b1}$ and the issuer's risk coordinates $x_{q2}$, moves with velocity $\boldsymbol{v}(t,\boldsymbol{x}_q)$ (A.2) in the economic domain (3.2). The particular trade with coordinates $\boldsymbol{x}_q$ at moment $t$ with velocity $\boldsymbol{v}(t,\boldsymbol{x}_q)$ carries its $m$-th trade value $C(\tau_k,\boldsymbol{x}_q;m)$ (3.10). The motion of numerous trades, which carry trade values, in the neighborhood of point $\boldsymbol{x}$ defines the *collective* flow and the *collective* velocity (A.3) of the *collective* trade value. One can define the *collective* flow $\boldsymbol{P}_C(\tau_k,\boldsymbol{x};m)$ and the *collective* velocity $\boldsymbol{v}_C(\tau_k,\boldsymbol{x};m)$ (A.3) of the *collective* trade value $C(\tau_k,\boldsymbol{x};m)$ (3.8) at point $\boldsymbol{x}$:

$$P_{\mathrm{C}}(\tau_k,\boldsymbol{x};m)=\sum_{\boldsymbol{x}_q\in dV(x)}C\left(\tau_k,\boldsymbol{x}_q;m\right)\boldsymbol{v}\left(t,\boldsymbol{x}_q\right)=C(\tau_k,\boldsymbol{x};m)v_{\mathrm{C}}(\tau_k,\boldsymbol{x};m) \qquad \text{(A.3)}$$

Let us consider the change of $C(t,\boldsymbol{x};m)$ in a small volume $\delta V(\boldsymbol{x})$ during the time $dt$. Two factors determine its change in a small volume $\delta V(\boldsymbol{x})$ (Childress, 2009). The first determines the change in time:

$$\delta V(\boldsymbol{x})dt\ \frac{\partial}{\partial t}C(t,\boldsymbol{x};m)$$

The second factor determines the change of $C(t,\boldsymbol{x};m)$ due to the in- and out- flows of $P_C(t,\boldsymbol{x};m)$ (A.3) in the small volume $\delta V(\boldsymbol{x})$. Indeed, the velocity $v_C(t,\boldsymbol{x};m)$ (A.2) carries in- and out- the amount of $C(t,\boldsymbol{x};m)$ through the borders of $\delta V(\boldsymbol{x})$ and that results in the total change of $C(t,\boldsymbol{x};m)$ inside $\delta V(\boldsymbol{x})$ during $dt$ as:

$$\delta X dt\frac{\partial}{\partial \boldsymbol{x}}\cdot\boldsymbol{P}_{\mathrm{C}}(t,\boldsymbol{x};m)$$

As $\delta X$ and $dt$ are arbitrary small, one obtains the equation of the total change of $C(t,\boldsymbol{x};m)$ (Childress, 2009):

$$\frac{\partial}{\partial t}C(t,\boldsymbol{x};m)+\frac{\partial}{\partial x}\cdot\boldsymbol{P}_{\mathrm{C}}(t,\boldsymbol{x};m)=\frac{\partial}{\partial t}C(t,\boldsymbol{x};m)+\frac{\partial}{\partial x}\cdot\left[C(t,\boldsymbol{x};m)\boldsymbol{v}_{\mathrm{C}}(t,\boldsymbol{x};m)\right]=F_C(t,\boldsymbol{x};m) \qquad \text{(A.4)}$$

The symbol "·" denotes scalar product and $\frac{\partial}{\partial x}\cdot\boldsymbol{P}_{\mathrm{C}}$ denotes the divergence. To derive the equations on the flow $\boldsymbol{P}_C(t,\boldsymbol{x};m)$ (A.3) repeat the procedure (Childress, 2009):

$$\frac{\partial}{\partial t}\boldsymbol{P}_{\mathrm{C}}(t,\boldsymbol{x};m)+\frac{\partial}{\partial x}\cdot\left[P_{\mathrm{C}}(t,\boldsymbol{x};m)\boldsymbol{v}_{\mathrm{C}}(t,\boldsymbol{x};m)\right]=\boldsymbol{G}_C(t,\boldsymbol{x};m) \qquad \text{(A.5)}$$

The factors $F_C(t,\boldsymbol{x};m)$ and $\boldsymbol{G}_C(t,\boldsymbol{x};m)$ in the right hand of (A.4; A.5) determine the impact of the economic environment on the *collective* trade values $C(t,\boldsymbol{x};m)$ (4.10) and their flows $\boldsymbol{P}_C(t,\boldsymbol{x};m)$ (A.3). These factors determine the economic origin of the market trade evolution. We call (A.4; A.5) equations of the continuous economic media approximation. The left sides of (A.4; A.5) have a common form of the continuous media equations and have been in use in textbooks (Childress, 2009) for a century. The economic origin of the model is completely different from the equations of physical hydrodynamics. The right side factors (A.4; A.5) describe the economic and market nature of the continuous economic media approximation. Some simple cases were described in Olkhov (2018-2020).

***Equations of motion of the whole market***

To describe the economic variables of the whole market, take the integrals of (A.4; A.5) by $d\boldsymbol{x}$ over the economic domain (3.2) and get ordinary differential equations:

$$\frac{\partial}{\partial t}C(t;m)=F_C(t;m)\quad ;\quad \frac{\partial}{\partial t}\boldsymbol{P}_{\mathrm{C}}(t;m)=\boldsymbol{G}_C(t;m) \qquad \text{(A.6)}$$

$$C(t;m) = \int C(t, \boldsymbol{x}_m; m)\, d\boldsymbol{x} \quad ; \quad \boldsymbol{P}_C(t;m) = \int \boldsymbol{P}_C(t, \boldsymbol{x}; m)\, d\boldsymbol{x} \qquad (A.7)$$

The total trade value $C(t;m)$ (A.7) of the whole market coincides with $C(\mu_k;m)$ (5.5) for $\mu_k=t$. Equations (A.6) have a simple form, but their complexities are hidden in the right-hand factors. The main results of the transition from equations (A.4; A.5) to equations (A.6) of the whole stock market are tied up with the hidden economic variables that significantly impact economic evolution. As an example, we highlight the mean risks linked to each economic variable. Let us consider the mean risk $\boldsymbol{X}_C(t;m)$ determined as:

$$\boldsymbol{X}_C(t;m)\, C(t;m) = \int \boldsymbol{x}\, C(t, \boldsymbol{x}; m)\, d\boldsymbol{x}$$

The vector $\boldsymbol{X}_C(t;m)$ determines the mean risk of the *joint* trade value $C(t;m)$ (5.5; A.7) in the economic domain (4.1). The components of $\boldsymbol{X}_C(t;m)$ fluctuate with time in the square $[0,1]^2$ of the economic domain (4.1). The change of the mean risk $\boldsymbol{X}_C(t;m)$ of the *joint* trade values $C(t;m)$ of the whole market is a slow process, and its fluctuations describe the cycles of the *joint* trade value $C(t;m)$ (5.5; A.7) that are alike to the business or credit cycles (Olkhov, 2020). The mean risk $\boldsymbol{X}_C(t;m)$ of the *joint* trade value $C(t;m)$ differs from the mean risk $\boldsymbol{X}_U(t;m)$ of the *joint* trade volume $U(t;m)$ (5.6) or mean risks linked with other collective economic variables. The hidden dynamics of mean risks describe the important properties of market trade evolution that are almost completely missed by current economic models.